\documentclass[pra,aps,showpacs,superscriptaddress,twocolumn,floatfix]{revtex4}
\usepackage{graphicx}
\usepackage[ansinew]{inputenc}
\usepackage{array}
\usepackage{color}
\usepackage{amsmath}
\usepackage{amsxtra}
\usepackage{amstext}
\usepackage{amssymb}
\usepackage{latexsym}
\usepackage{dsfont}
\usepackage{verbatim}
\usepackage{comment}
\begin{document}

\title{Spinor atom-molecule conversion via laser-induced three-body recombination}
\author{H. Jing$^{1,2}$, Y. Deng$^{1}$, and P. Meystre$^{2}$}
\affiliation{$^1$Department of Physics, Henan Normal University,
Xinxiang 453007, China\\
$^2$B2 Institute, Department of Physics and College of Optical
Sciences, The University of Arizona, Tucson, Arizona 85721}
\date{\today}

\begin{abstract}
We study theoretically several aspects of the dynamics of coherent
atom-molecule conversion in spin-1 Bose-Einstein condensates.
Specifically, we discuss how for a suitable dark-state condition the
interplay of spin-exchange collisions and photoassociation leads to
the stable creation of an atom-molecule pairs from three initial
spin-zero atoms. This process involves $two$ two-body interactions
and can be intuitively viewed as an effective three-body
recombination. We investigate the relative roles of photoassociation
and of the initial magnetization in the ``resonant'' case where the
dark state condition is perfectly satisfied. We also consider the
"non-resonant" regime, where that condition is satisfied either
approximately -- the so-called adiabatic case -- or not at all. In
the adiabatic case, we derive an effective non-rigid pendulum model
that allows one to conveniently discuss the onset of an
antiferromagnetic instability of an ``atom-molecule pendulum,'' as
well as large-amplitude pair oscillations and atom-molecule
entanglement.
\end{abstract}

\pacs{42.50.-p, 03.75.Pp, 03.70.+k} \maketitle

\section{Introduction}

Recent years have witnessed rapid advances in the manipulation of
the spin degrees of freedom of ultracold atoms \cite{Meystre, spin,
spin 1, spin domains, spin-2-2, Cr}. By magnetically steering
two-body collisions, a broad range of effects has been observed,
including atomic magnetism \cite{Ho98,Ohmi98,Law98, Pu99}, coherent
spin mixing \cite{spin mixing,spin-2-2}, topological excitations
\cite{votex}, and an atomic analog of the Einstein-de Haas effect
\cite{de Haas}. The {\em optical} control of atomic spin dynamics
has also attracted much experimental interest
\cite{Dum,Chapman,APB}. For example, Dumke {\em et al.} \cite{Dum}
and Hamley {\em et al.} \cite{Chapman} have investigated the
photoassociation (PA) diagnosis \cite{Dum} and PA spectroscopy
\cite{Chapman} of spin-1 atoms, opening the way to studies of
PA-controlled regular \cite{HJ} or chaotic \cite{J. C.} spin
dynamics.

In a very recent experiment, the ro-vibrational ground-state
molecules were successfully prepared via the all-optical association
of laser-cooled atoms \cite{Inouye}, which has triggered the
investigation of coherent PA of a wide variety of ultracold atomic
and molecular systems~\cite{Carr}. A result of particular relevance
for the present study is an experiment by Kobayashi {\em et al.},
who used a coherent two-color PA technique to create spinor
molecules in a spin-1 atomic Bose condensate \cite{APB}. In
particular, these authors found that for strong PA couplings the
atomic spin oscillations are significantly suppressed and the
dominant process is scalar-like atom-molecule conversion. That is,
only the populations of the spin components that are associated into
molecules are observed to decrease, while the other spin component
remains almost unchanged on the experimentally relevant timescale
\cite{APB}.

In this paper we show that under appropriate two-photon resonance conditions quantum interferences between optical PA and atomic spin mixing can lead to the existence of a dark state of the spin-down atoms, which can in turn be exploited in the stable formation of a spinor atom-molecule pair from three initial spin-zero atoms. This process, which involves $two$ two-body interactions, can be thought of as an effective three-body spin-exchange effect. The important role of the initial magnetization in creating the atom-molecule pairs is also analyzed. We also analyze dynamical features that occur in the ``non-resonant'' regime where no dark state is formed, including large-amplitude coherent oscillations of the atom-molecule pairs population and an antiferromagnetic instability. As such, these manifestations of the interplay between two-color PA and spin-exchange collisions sheds significant new insight into the study of quantum spin gases and ultracold chemistry \cite{Carr}.

The article is organized as follows. Section II discusses the
"resonant" situation where the dynamics of the system is
characterized by the existence of a dark state. We first introduce
our model, which we then apply to the description of scalar-like
photoassociation~\cite{APB}. We then derive a dark-state condition
for the spin-down atoms and show that when satisfied, it results in
the stable resonant creation of atom-molecule pairs. The role of the
initial atomic magnetization is also discussed. Section III then
turns to the non-resonant regime. We show that in that case the
system can be described in terms of a nonrigid pendulum model. Two
important dynamical manifestations of this regime, large-amplitude
atom-molecule oscillations, and a regime of antiferromagnetic
instability are explicitly discussed. Finally Section IV is a
summary and conclusion.

\begin{figure}[tbp]
\includegraphics[width=2.5in]{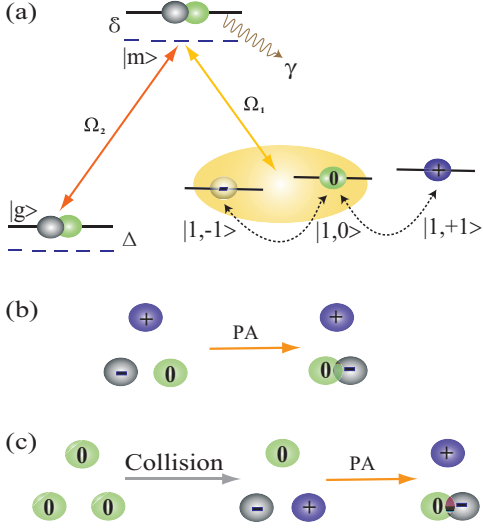}
\caption{(Color online). (a) Schematic of coherent two-color PA in a spin-1 atomic condensate. Here $\delta$ and $\Delta$ are the one-  and two-photon detunings of the laser fields with Rabi frequencies $\Omega _{1,2}(t)$, and $\gamma $ accounts for the spontaneous decay of the excited state $|m\rangle $. (b) Scalar-like atom-molecule conversion as observed in a recent experiment of Kobayashi {\em et al.} \cite{APB}. (c) Effective three-body recombination resulting from the interplay of $two$ two-body interactions (see text). } \label{fig1}
\end{figure}

\section{The model}

This section introduces our model and  exploit it to describe the
main features of scalar-like PA~\cite{APB}. We also discuss a regime
of stable atom-molecule pair formation, and analyze the role of
initial magnetization in the system dynamics.

\subsection{Theoretical model}

The system that we consider is illustrated in Fig.~1. It consists of a spin-1 atomic condensate undergoing spin-changing two-body collisions and coupled via 2-photon coherent PA to a ground-state
diatomic molecular condensate.

Denoting by $\hat{\psi}_{i,j=0,\pm 1}$ and $\hat{\psi}_{m,g}$ the
annihilation operators of the three atomic components and of the
excited or ground-state molecules, respectively, the Hamiltonian of
the binary atomic and molecular condensate is $(\hbar =1)$
\begin{equation}
\hat{{H}}=\hat{\mathcal H}_0 +  \hat{\mathcal{H}}_{c\rm oll}+\hat{\mathcal{H}}_{\rm PA},
\end{equation}
where
\begin{eqnarray}
\hat{\mathcal{H}}_0&=&\int d{\bf r}\left [ \sum_{i=-1, 0, 1}\hat{\psi}_i^{\dag }\left (V+E_i \right )\hat{\psi}_i  \right . \nonumber \\
&+& \left . \left (\delta -\frac{1}{2}i\gamma \right )\hat{\psi}_m^{\dag }\hat{\psi}_m + (\Delta+\delta)\hat{\psi}_{g}^{\dag }\hat{\psi}_{g} \right ],\\
\hat{\mathcal{H}}_{\rm coll}&=& \frac{1}{2}\int d\bf{r} \left  [ c_0^\prime \hat{\psi}_i^\dag \hat{\psi}_j^\dag \hat{\psi}_j \hat{\psi}_i  \right . \nonumber \\
&+& \left .c_2^\prime \hat{\psi}_i^\dag(F_\kappa)_{ij}\hat{\psi}_j \hat{ \psi}_k^\dag (F_\kappa)_{kl}\hat{\psi}_l \right ], \\
\hat{\mathcal{H}}_{\rm PA}&=& \int d {\bf r} \left [ -\Omega _2\hat{\psi}_g^\dag \hat{ \psi}_m +\Omega_1 \hat{\psi}_m^\dag \hat{\psi}_0\hat{\psi}_{-1}+H.c.  \right ].
\end{eqnarray}
Here  $V$ is the trap potential, $E_i$ is the energy of the spin
state $i$ with a static magnetic field lifting their degeneracy,
$F_{\kappa =x,y,z}$ are spin-1 matrices, and
$$
c_0'=4\pi(a_0+2a_2)/3M
$$
and
$$
c_2'=4\pi(a_2-a_0)/3M
$$
where  $a_{0,2}$ are $s$-wave scattering lengths \cite{Ho98}. Finally  $\Omega_i, i=\{1,2\}$ are the Rabi frequencies of the PA fields, and $\gamma$ is a phenomenological decay factor. The detunings $\delta$ and $\Delta$ between the PA fields and the atomic and molecular levels are defined in Fig.~1. We have ignored the kinetic energy of the particles by assuming a dilute and homogeneous ensemble. Note also that this model ignores collisions between the molecules since there is currently no knowledge of their strength.

To extract the main aspects of the system dynamics we invoke a single-mode approximation, a simplification that has proven successful in describing key aspects of related systems in the past
\cite{spin, Law98, Pu99}. It amounts to approximating the fields operators of the three spin components of the atomic condensate as
$$
\hat{\psi}_i(\vec r,t)=\sqrt N \hat a_i(t)\phi(\vec r)\exp(-i\mu
t/\hbar),
$$
where $N$ is the initial atomic number, $\mu$ the chemical potential, $\phi(\vec r)$ is the normalized condensate wave function for each spin component, satisfying $\hat{\mathcal{H}}_{S}\phi(\vec r)=\mu\phi(\vec r)$ with $\int d \vec r|\phi(\vec r)|^2=1$, and $\hat{a}_i(t)$ are bosonic annihilation operators. The molecular condensate is described likewise in a single-mode approximation, with the annihilation operators $\hat m$ and $\hat g$ describing excited and ground-sate molecules.

For large enough detunings $\delta$ the intermediate molecular state $|m\rangle$ can be adiabatically eliminated \cite{adiabatic elimination}, simplifying the Heisenberg equations of motion of the atom-molecule system to
\begin{eqnarray}\label{heisenberg}
i\frac{d\hat{a}_{+}}{d\tau}&=& \chi_2(\rho_+ + \rho_0 -\rho_-)\hat{a}_{+}+\chi_2\hat{a}_{0}^{2}\hat{a}_{-}^{\dagger},\nonumber\\
i\frac{d\hat{a}_{0}}{d\tau}&=& \chi_2(\rho_+ +\rho_-)\hat{a}_{0}-\omega\rho_-\hat{a}_0+2\chi_2\hat{a}_{+}\hat{a}_{-}\hat{a}_{0}^{\dagger} +\Omega\hat{g}\hat{a}_-^{\dagger},\nonumber\\
i\frac{d\hat{a}_{-}}{d\tau}&=& -\Gamma\hat{a}_-+\chi_2\hat{a}_{0}^{2}\hat{a}_{+}^{\dagger} +\Omega\hat{g}\hat{a}_0^{\dagger},\nonumber\\
i\frac{d\hat{g}}{d\tau}&=& \Omega\hat{a}_{0}\hat{a}_{-}+(\Delta+\delta-\delta')\hat{g},
\end{eqnarray}
where
\begin{eqnarray}
c_{0,2}&=&c_{0,2}^\prime \int d\mathbf{r}|\phi(\mathbf{r)|^4},\\
\delta^{\prime}&=&\frac{\Omega_2^2}{c_0N\delta}\left(1+\frac{i\gamma}{2\delta}\right)
\end{eqnarray}
and we have introduced the dimensionless variables $\tau=c_0Nt$, $\chi_2=c_2/c_0$, $\omega= \Omega_1^2/(c_0N\delta)$, $\Gamma=\omega\rho_0-\chi_2(\rho_- + \rho_0 - \rho_+)$, and
$$
\Omega=\frac{\Omega_1\Omega_2}{c_0N\delta}.
$$

\subsection{Scalar-like photoassociation}

In their recent experiment on two-color PA of the spinor atoms $^{87}$Rb \cite{APB}, Kobayashi {\em et al.} observed the spin-selective formation of the molecular state $|2, -1\rangle$ from reactant atoms in the state $|1, -1\rangle$ and $|1, 0\rangle$. One important feature of their experimental results is that while the populations of the reactant atoms decreased, the population of the state $|1, 1\rangle$ remained almost unchanged. This is the situation  illustrated in Fig.~1(b) \cite{APB}.

To test our model against that experiment we assume that the energy degeneracy of the atomic magnetic sublevels is lifted by
 a static magnetic field and that the atomic condensate is initially prepared in the state $f=[\sqrt{0.2}, \sqrt{0.6}, \sqrt{0.2}]$ \cite{APB}.
 The experiment used two lasers of maximum powers $I_1=I_2/2= 10 W$, detuning $\delta= 2\pi \times 300$MHz and $\Omega/\sqrt{I}=7 {\rm MHz(W cm}^{-2})^{-\frac{1}{2}}$,
 which yields in our case $\Omega_1=139$ MHz and $\Omega_2=$197 MHz. As we will see in the following these values are well beyond the regime of atom-molecule pair formation,
 and as illustrated in Fig.~2 our model does confirm that the two-color PA of atoms into molecules is scalar-like in this case.

\begin{figure}[tbp]
\includegraphics[width=2.75in]{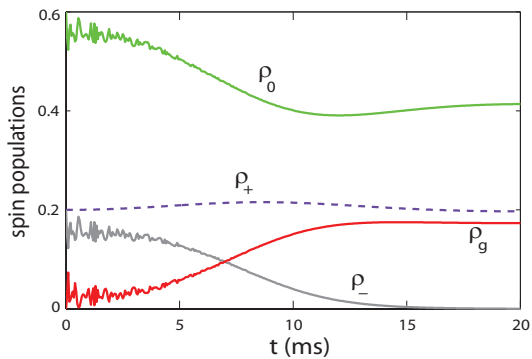} \center \caption{(Color online)
Scalar-like atom-molecule conversion of $^{87}$Rb atoms, with essentially unchanged population of the spin-up state~\cite{APB}. The initial condition is $f=[\sqrt{0.2}, \sqrt{0.6}, \sqrt{0.2}]$, and $\Omega=\Omega_m {\rm sech}(t/4)$ with $|{\Omega_m}/{\chi_2}|=1.44\times 10^4$ \cite{APB}. The other parameters are $\chi_2=-0.01$, $\delta=-100\chi_2$, $\gamma=10|\chi_2|$, and $c_0N=10^5s^{-1}$.}
\end{figure}

\subsection{Stable atom-molecule pair formation}

The scalar-like photoassociation sketched in the previous subsection
results from the binding of a pair of Rb atoms of spin-$0$ and
spin-down. We now consider the case of PA from spin-0, but in the
presence of spin-changing collisions, the situation sketched in
Fig.~1(c). Specifically, we assume that the atomic condensate is
initially prepared in the spin-$0$ state $|1, 0\rangle$.
Spin-exchange collisions couple then a pair of spin-0 atoms to a
pair of atoms with opposite spins, $2A_0\rightarrow A_\downarrow
+A_\uparrow$ \cite{Chapman}, while PA fields of appropriate
wavelengths selectively combine a spin-down atom and a spin-0 atom
into the molecular ground state $|g\rangle$ via a virtual transition
to an excited molecular $|m\rangle$, $A_0+A_\downarrow\rightarrow
A_0A_{\downarrow}$ \cite{APB}. The outcome of these combined
mechanisms is the creation of an atom-molecule pair from three
spin-0 atoms, $3A_0\rightarrow A_0A_{\downarrow}+A_\uparrow$, a
process that can be intuitively thought of as an effective,
spin-dependent three-body recombination. As such, this process is
quite different from both the scalar-like PA of the previous
subsection \cite{APB} and the purely atomic laser-catalyzed spin
mixing \cite{HJ}.

We found numerically that in this case the stable atom-molecule pair
formation is possible, provided that the dark-state condition
\begin{equation}
\Omega(t)=-\chi_2\sqrt{\frac{\rho_0\rho_+}{\rho_g}},
\label{dark state}
\end{equation}
for the spin-down atomic state is satisfied \cite{Ling,dark state}.
This result is easily confirmed from Eqs.~(\ref{heisenberg}), which
show that when condition~(\ref{dark state}) is satisfied the
spin-down atomic state remains essentially unoccupied. That
situation is illustrated in Fig.~3, which shows the efficient stable
creation of atom-molecule pairs in this case.

\begin{figure}[tbp]
\includegraphics[width=2.75in]{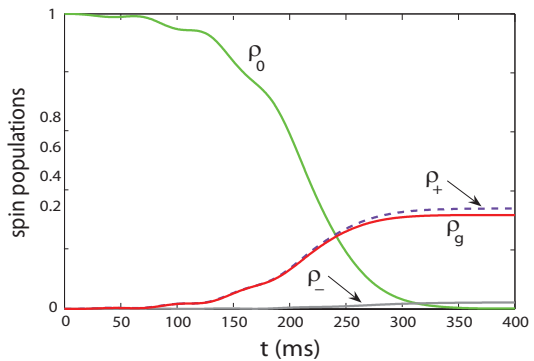} \center \caption{(Color online)
Atom-molecule pairs formation as a function of time for $^{87}$Rb
atoms, under the dark state condition for spin-down atoms. The
initial state is $f=[0, 1, 0]$ and the other parameters are the same
as Fig. 2.}
\end{figure}

\subsection{Role of magnetization}

The initial magnetization
\begin{equation}
\mathcal{M}=\rho_+-\rho_--\rho_g
\end{equation}
of a spin gas prepared in the state $f=[0, 1, 0]$ is
$\mathcal{M}=0$. In this subsection we consider the role of that
initial magnetization in the creation of atom-molecule pairs. We
find that in contrast to the case of scalar-like molecule formation,
the initial magnetization now plays a significant role, as
illustrated in Figs.~4 and 5.

Figure~4 shows the evolution of the population of ground-state
molecules  for several values of the initial magnetization, under
the generalized dark-state condition (\ref{dark state}). For
$\mathcal{M}\geq 0$ the ground-state molecules are produced
efficiently and reach a steady-state population
$\rho_g=(1-\mathcal{M})/3$; for $\mathcal{M}<0$, in contrast, this
population exhibits large oscillations -- see also Fig.~5, which
shows more details of the oscillations of $\rho_g$ for negative
magnetizations --  and do not appear to reach a steady state. This
is due to the simple fact that for $\mathcal{M}<0$, the populations
of spin-down atomic state are not zero and thus that state is a
"bright" state that does not remain uncoupled to the other atomic
states during association.

\begin{figure}[tbp]
\includegraphics[width=2.45in]{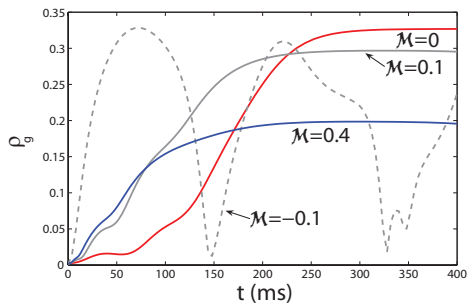} \center
\caption{(Color online) Spinor molecules population for several
values of the initial magnetization $\mathcal{M}$ under the
dark-state condition, with $\chi_2=-0.01$ and $\delta=100|\chi_2|$.
The stable formation of spinor molecules is possible only for
$\mathcal{M}\geq0$.}
\end{figure}

\begin{figure}[tbp]
\includegraphics[width=2.45in]{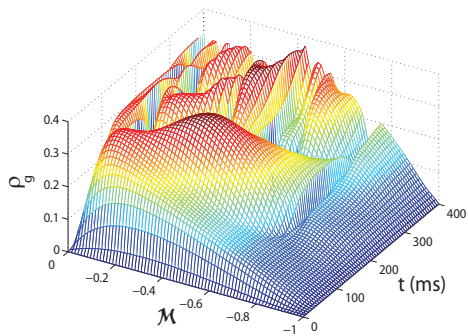} \center
\caption{(Color online) Large-amplitude oscillations of the spinor
molecules population for negative values of the magnetization
$\mathcal{M}<0$. All other parameters are as in Fig. 4.}
\end{figure}

\section{Non-resonant regime}

The dynamics of atom-molecule pair formation in the case of negative
magnetization indicates that the presence or absence of an atomic
dark state plays a key role in that process. In this section we
further investigate the ``non-resonant'' situation where no dark
state exists. We consider specifically two examples:  The first one
is an `adiabatic' case characterized by an approximate dark-state
condition. In this case the system dynamics can be understood in
terms of an effective nonrigid pendulum model that permits to
discuss an antiferromagnetic instability of the atom-molecule
pendulum. In a second example, we briefly discuss a situation where
the dark-state condition is strongly violated.

\subsection{Adiabatic case}

Figure~6 shows an example of atom-molecule-pair oscillations for a
non-resonant situation and starting from spin-0 $^{87}$Rb atoms.
(Note that pair formation implies that $\rho_+\simeq\rho_g$.) As
would be intuitively expected, the numerical integration of
Eqs.~(\ref{heisenberg}) confirms that the creation of atom-molecule
pairs is only possible for PA field strengths that allow for the
simultaneous occurrence of spin-exchange collisions and
atom-molecule conversion. For the initial atomic state $f = [0, 1,
0]$, we find that the Rabi frequencies of the PA fields should be
such that
$$
\Omega=-\chi_2\sqrt{\rho_0}\leq |\chi_2|
$$
 or equivalently
$$
\Omega_1\Omega_2 \leq |N\delta c_2|,
$$
which gives $\Omega_1\leq 0.3\pi$MHz, and $\Omega_2\leq 0.6\pi$MHz for the case $\Omega_2/\Omega_1$ = 2 considered here.

\begin{figure}[tbp]
\includegraphics[width=2.75in]{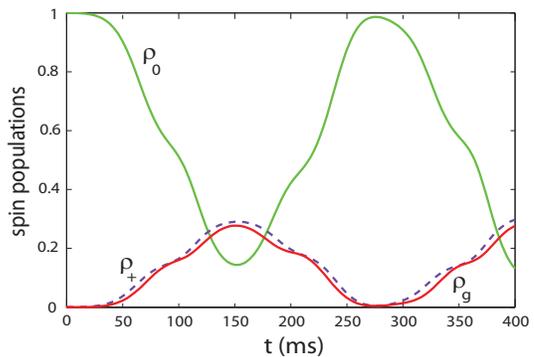} \center \caption{(Color online)
Coherent atom-molecule oscillations as a function of time for $^{87}$Rb atoms. The dashed line is the population of the spin-up atoms. The initial atomic state is $f=[0, 1, 0]$, $\Omega= 0.75 |\chi_2|$, and
the other parameters are as in Fig.~2. }
\end{figure}

We remark that for an atomic condensate initially prepared in the spin-0 state, and assuming that the dark-state condition (\ref{dark state}) is approximately satisfied, the first derivatives of the slowly-varying amplitudes for spin-down atoms can be neglected, $i\dot{\hat{a}}_{-}\approx 0$ \cite{Pu2000,adiabatic elimination 1}. It is then possible to describe the system by the approximate effective three-state
Hamiltonian
\begin{eqnarray}\label{H3}
\hat{\mathcal{H}}_{\rm eff} &=& \chi_3(\hat{a}_0^{\dagger 3}\hat{a}_+ \hat{g}+\hat{a}_0^3\hat{a}_+^\dagger \hat{g}^\dagger  \nonumber \\
&+& \frac{1}{\Gamma}(\Omega^2 \hat{\rho}_0 \hat{\rho}_g + \chi_2^2
\hat{\rho}_0^2\hat{\rho}_+) + \chi_2 \hat{\rho}_0 \hat{\rho}_+,
\end{eqnarray}
where $\chi_3={\Omega \chi_2}/{\Gamma }$.
The first term in this Hamiltonian describes the creation of atom-molecule pairs from three spin-0 atoms through a laser-induced effective three-body recombination \cite{three body}.

For short enough times, it is possible to neglect the depletion of the spin-0 population and to treat $\hat a_0$ as a c-number, $\hat a_0 \rightarrow N^{1/2}$. Linearizing the Hamiltonian (\ref{H3}), the second line reduces then to a simple self-interacting contribution, and the Heisenberg equations of motion for the remaining operators $\hat a_+$ and $\hat g$ have the solution
\begin{eqnarray}
\hat{a}_+(t) &=&\hat{a}_+(0)\cosh \chi'_3 t- i\hat{g}^{\dag }(0)\sinh\chi'_3 t,\nonumber\\
\hat{g}(t) &=& \hat{g}(0)\cosh\chi'_3 t-i\hat{a}_+^{\dag}(0)\sinh\chi'_3 t,
\end{eqnarray}
with $\chi'_3=N^{3/2}\chi_3$. These solutions are well-known to be indicative of quantum entanglement of the created atom-molecule pairs. As such this system is formally a matter-wave analog of optical parametric down conversion in quantum optics \cite{Meystre,Pu2000}.

\subsection{Antiferromagnetic instability}

Within the mean-field approach, the spatial part of the atomic and molecular wave functions can be written as  $\sqrt{N}e^{-i\mu t/\hbar}\zeta$, where $ \zeta \sqrt{\rho_i}e^{i\theta_i}$ or $\sqrt{\rho_g }e^{i\theta_g}$ and $\theta_i$ represents the phase of the $i$-th Zeeman state \cite{spin}. Within this description the dynamics of the system can be expressed in terms of the coupled equations
\begin{eqnarray}
\dot \rho_0 &=&{3 \chi_3}{\rho_0^{3/2}} \sqrt{(1-\rho_0)^2-\mathcal{M}^2} \sin \theta, \nonumber\\
\dot \theta &=&-{\Theta} +\chi_2(1+\mathcal{M}-2\rho_0) +{\frac{1}{\Gamma}} [ \chi_2^2\rho_0(3+3\mathcal{M}-4\rho_0)  \nonumber \\
&+&\Omega ^2({\frac{3}{2}} - {\frac{3m}{2}} -{\frac{5\rho_0}{2}}) + \Omega^2+(\Delta+\delta-\delta')\Gamma]\nonumber \\
&+&{\frac{\Omega \chi_2}{2\Gamma}}{\frac{\sqrt{\rho_0}[(1-\rho_0)(9-13 \rho_0)-9\mathcal{M}^2] }{\sqrt{(1-\rho_0)^2-\mathcal{M}^2}}} \cos\theta,
\label{canonical}
\end{eqnarray}
where
\begin{equation}
\theta =3\theta_0 - (\theta_+ +\theta_g)
\end{equation}
 and
\begin{equation}
\Theta=E_g+E_+-3E_0.
\end{equation}
These nonlinear equations support the two phase-independent fixed-point solutions $\rho_0=0$ and $\rho_0=1-|\mathcal{M}|$, as well as phase-dependent solutions for $\theta=0$ or $\pi$.

Equations~(\ref{canonical}) describe a nonrigid pendulum with energy functional
\begin{equation}
\label{E}
\mathcal{E}=\lambda_1 \cos \theta +\lambda_2,
\end{equation}
where
\begin{eqnarray}
 \label{energy}
\lambda_1&=&{3 \chi_3}{\rho_0^{3/2}} \sqrt{(1-\rho_0)^2-\mathcal{M}^2},\nonumber \\
\lambda_2&=&\frac{\rho_0}{\Gamma}\left [\chi_2^2\rho_0\left (\frac{3}{2} +\frac{3}{2}\mathcal{M}-\frac{4}{3}\rho_0\right ) \right . \nonumber \\ &+& \left . \frac{\Omega^2}{2}\left (3-3\mathcal{M}-\frac{5}{2}\rho_0\right )\right ]  \nonumber \\
&-& \rho_0\left (\Theta +\frac{\Omega^2}{\Gamma}+\Delta+\delta -  \delta^\prime\right ) \nonumber \\
&+&\chi_2\rho_0(1+\mathcal{M}-\rho_0).
\end{eqnarray}
This approach allows one to study simply the stability of the
magnetic domain structure of the system. Specifically, we follow the
approach of Ref.~\cite{Zhang} and consider instabilities associated
with a change in the sign of $d{\mathcal E}/{d{\cal M}}$. For
example, $dE/d{\cal M} > 0$ for ${\cal M}>0$ and $dE/d{\cal M} < 0$
for ${\cal M}<0$ implies that the magnetization always oscillates
around zero, and no domain forms. Following this approach we find
that, in contrast to the situation for purely atomic gases
\cite{Sadler,Zhang}, an instability of the domain structure can
occur for both ferromagnetic and anti-ferromagnetic atoms.  One
finds readily from Eqs.~(\ref{E}) and (\ref{energy}),
\begin{eqnarray}
\frac{d \cal E}{d {\cal M}}&=&\frac{3\chi_3}{2}{\cal M}
\left[1-\frac{\rho_0^{3/2}cos\theta}{\sqrt{(1-\rho_0)^2-{\cal M}^2}}\right]
+\chi_2\rho_0 \nonumber\\
&+&\frac{3\rho_o}{2\Gamma}(\chi_2^2\rho_0-\Omega^2).
\end{eqnarray}

\begin{figure}[tbp]
\includegraphics[width=2.55in]{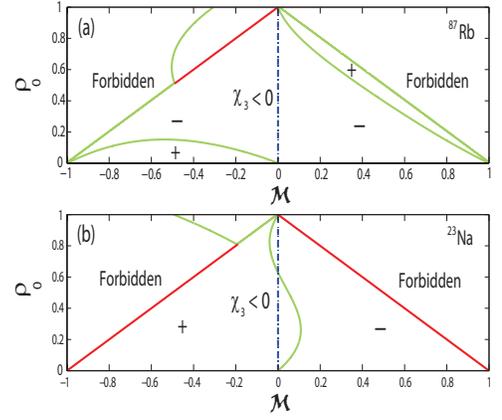} \center
\caption{(Color online)  Surfaces of $d \mathcal{E}$/$d
\mathcal{M}=0$ (green solid lines) for (a) ferromagnetic $^{87}$Rb
atoms ($\theta=0$, $\chi_2=-0.01$); and (b) anti-ferromagnetic
$^{23}$Na atoms ($\theta=\pi$, $\chi_2=0.01$). The red forbidden
line is determined by the condition of conserved total atomic number
or $\rho_0 + |\cal{M}|$ $\leq 1$ (see Ref. \cite{Zhang}).}
\end{figure}
Figure~7 shows the resulting surfaces of $d \mathcal{E}$/$d {\cal
M}=0$ for the ferromagnetic and anti-ferromagnetic cases. The plus
or minus sign denotes $d \mathcal{E}$/$d \mathcal{M}>0$ or $d
\mathcal{E}$/$d \mathcal{M}<0$. Here the condensate size is already
assumed to be much larger than the healing length $\mathcal
{L}_s=2\pi/\sqrt{2M|c_2'|n}$ at least in one direction so that
instability-induced domains can appear \cite{Zhang}.

As already mentioned, for $d\mathcal{E}$/$d \mathcal{M} < 0$ an
increase in $\mathcal{M}$ leads to lower energy while for $d
\mathcal{E}$/$d \mathcal{M}>0$ it leads to a higher energy. Hence
the (+, -) boundary delimitates the domain of dynamic instability
(see e.g. Ref. \cite{Zhang} for more details). We observe that in
contrast to the case of a pure sample of $^{87}$Rb atoms, which is
characterized by a wide instability region~\cite{Zhang}, in the case
at hand this region can be significantly reduced by an appropriate
tuning of the lasers. We also note that in the case of
anti-ferromagnetic atoms such as $^{23}$Na, where no dynamical
instability exists for a pure atomic sample, for our hybrid system,
an instability can now develop for a wide range of parameters, see
Fig.~7b.

One point to emphasize is that the antiferromagnetic instability can be experimentally observed without any laser fields, i.e. for $\Omega=0$ -- although these fields are of course required for the formation of molecules. We also remark that the spin mixing of spin-2 molecules is slow enough in comparison with the effective three-body recombination process that it can be safely ignored here. However, thermalization and spontaneous decay of the ground-state molecules are expected to be major challenges for the observation of coherent oscillations of atom-molecule pairs \cite{spin-2-2}.

\subsection{Violation of the dark-state condition}

As a final special case we now consider the situation when
$|\Omega/\chi_2|>1$, in which case the dark-state condition
(\ref{dark state}) is completely violated. Figure~8 shows that for
increasing values of $\Omega/\chi_2$, the amplitude of the
oscillations in molecular population first increase, and then
decreases until  $|\Omega/\chi_2| =1$. Beyond that critical value
the molecular oscillations become strongly damped, and eventually
population transfer to the molecular ground state essentially
disappears, as illustrated in the figure for $|\Omega/\chi_2=1.5$.
As illustrated in Fig.~8(b) the population oscillations of spin-$0$
atoms is also strongly suppressed in that regime of strong PA.

Finally Fig.~8(b) also illustrates how different choices of the
initial atomic state result in different dynamics of the spinor
atom-molecule system. In particular, an atomic sample initially in
the spin-0 state remains completely unperturbed by the strong PA
fields (far from the dark-sate resonance condition). Note that the
scalar-like atom-molecule conversion illustrated in Fig.~2
corresponds to fields that strongly violate the condition (\ref{dark
state}), with $|\Omega/\chi_2| = 1.44 \times 10^4 \gg 1$. In that
case the only parameters of practical relevance are the initial
atomic state and the strengths of the PA fields.
\begin{figure}[tbp]
\includegraphics[width=2.75in]{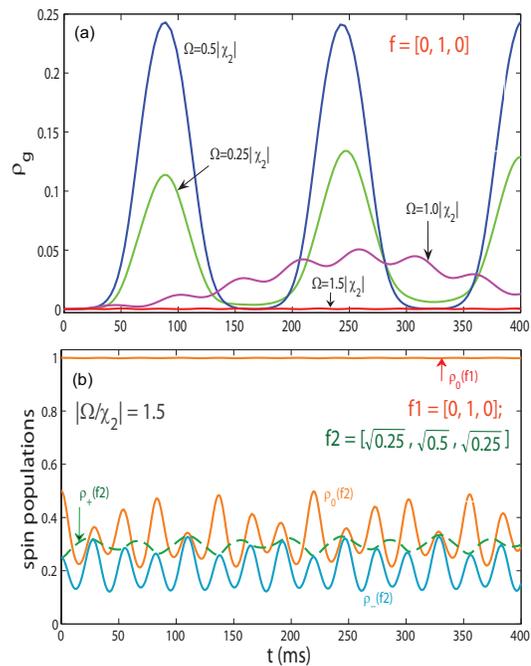} \center \caption{(Color online) (a) Molecular oscillations for several values of $|\Omega/\chi_2|$,
which label the curves, and  the initial atomic state $|0, 1,
0\rangle$. (b) Atomic spin populations for the initial atomic states
$|f_1\rangle = |0, 1,0\rangle$ and $|f_2\rangle = |\sqrt{0.25},
\sqrt{0.5}, \sqrt{0.25}\rangle$, and  for $|\Omega/\chi_2|=1.5$.
Other parameters are as in Fig.~ 2. }
\end{figure}

\section{Summary and Conclusion}

In conclusion, we have studied a number of aspects of coherent
photoassociation in a spinor Bose condensate, with emphasis on the
creation of atom-molecule pairs from the initial spin-zero atoms.
This process, which involves $two$ two-body interactions, can be
conveniently described by an effective three-body spin-dependent
recombination mechanism -- the term "three-body recombination" being
used here to differentiate our proposal from the recent two-color PA
experiment (that involves the scalar-like association of spinor
atoms) \cite{APB}. We have shown in particular that the spin-down
atoms can be kept in a dark state for appropriate conditions in both
the initial states of the atoms and PA fields, leading to the
formation of atom-molecule pairs. For comparison we also considered
the regimes with PA fields strong enough to violate the dark-state
condition.

Although it shares the similar usage of PA fields and spin-dependent
collisions, the present work is different from previous results on
laser-catalyzed atomic spin oscillations \cite{HJ}, which did not
involve the formation of molecules. In addition, the simulations of
experimentally observed scalar-like features in associating spinor
atoms, the study of the roles of magnetization and of the initial
atomic state, and the antiferromagnetic instability of a hybrid
atom-molecule system are also the new results.

In view of the rapid experimental advances in all-optical
association of laser-cooled atoms \cite{Inouye}, it can be expected
that the coherent PA of quantum spin gases, in particular, the
atom-molecule pair formation in a spinor sample, should become
experimentally observable in the near future~\cite{APB}.
Laser-controlled spinor reactions can provide a new testing ground
to address a number of questions in many-body physics, cold
chemistry, and quantum information science. Future work will study
the creation of heteronuclear spinor molecules from a two-species
atomic spin gas \cite{hetero}, and the spinor reactions in an
optical lattice \cite{Daley}, with and without the long-range
dipole-dipole interactions~\cite{de Haas}. We also plan to study the
cavity-assisted amplification of spinor molecules~\cite{CPA}, the
bistability of a spinor atom-molecule ``pendulum''~\cite{Ying}, and
the spinor trimer formation~\cite{Carr,trimer}.

This work is supported by the U.S. Office of Naval Research, by the U.S. National Science Foundation, by the U.S. Army Research Office, and by the National Science Foundation of China under Grant Numbers
10874041 and 10974045.


\begin{thebibliography}{99}
\bibitem{Meystre} P. Meystre, Atom Optics (Springer-Verlag, Berlin, 2001).

\bibitem{spin domains} J. Stenger, S. Inouye, D. M. Stamper-Kurn1, H.-J. Miesner, A. P. Chikkatur, and W.
Ketterle, Nature (London) \textbf{396}, 345 (1998).

\bibitem{spin} M.-S. Chang.1, Q. Qin, W. Zhang, L. You, and M. S. Chapman, Nature Phys. \textbf{1}, 111 (2005).

\bibitem{spin 1} M.-S. Chang, C. D. Hamley, M. D. Barrett, J. A. Sauer, K. M. Fortier, W. Zhang, L. You, and M. S. Chapman, Phys. Rev. Lett. \textbf{92}, 140403 (2004).

\bibitem{spin-2-2} H. Schmaljohann, M. Erhard, J. Kronj\"ager, M. Kottke, S. van Staa, L. Cacciapuoti, J. J. Arlt, K. Bongs, and K. Sengstock,
 Phys. Rev. Lett. \textbf{92}, 040402 (2004); J. Kronj\"{a}ger, C. Becker, P. Navez, K. Bongs, and K. Sengstock,
$ibid$. \textbf{97}, 110404 (2006).

\bibitem{Cr} A. Griesmaier, J. Werner, S. Hensler, J. Stuhler, and T. Pfau, Phys. Rev. Lett. \textbf{94}, 160401 (2005).

\bibitem{Ohmi98} T. Ohmi and K. Machida, {J. Phys. Soc. Jpn.} \textbf{67}, 1822 (1998).

\bibitem{Ho98} T.-L. Ho, {Phys. Rev. Lett.} \textbf{81}, 742 (1998).

\bibitem{Law98} C. K. Law, H. Pu, and N. P. Bigelow, {Phys. Rev. Lett.} \textbf{81}, 5257 (1998).

\bibitem{Pu99} H. Pu, C. K. Law, S. Raghavan, J. H. Eberly, and N. P.
Bigelow, Phys. Rev. A \textbf{60}, 1463 (1999); W. X. Zhang, D. L.
Zhou, M.-S. Chang, M. S. Chapman, and L. You, $ibid$ \textbf{72},
013602 (2005).

\bibitem{spin mixing} B. Sun, W. X. Zhang, S. Yi, M. S. Chapman, and L. You, {Phys. Rev. Lett.} \textbf{97}, 123201 (2006).

\bibitem{votex} A. E. Leanhardt, A. G\"orlitz, A. P. Chikkatur, D. Kielpinski, Y. Shin, D. E. Pritchard, and W. Ketterle,
Phys. Rev. Lett. \textbf{89}, 190403 (2002);  M. Vengalattore, S. R. Leslie, J. Guzman, and D. M. Stamper-Kurn,
$ibid$. \textbf{100}, 170403 (2008).

\bibitem{de Haas} Y. Kawaguchi, H. Saito, and M. Ueda, Phys. Rev. Lett. \textbf{96}, 080405 (2006).

\bibitem{Dum} R. Dumke, M. Johanning, E. Gomez, J. D. Weinstein, K. M. Jones, and P. D. Lett, New J. Phys. \textbf{8}, 64 (2006).

\bibitem{Chapman} C. D. Hamley, E. M. Bookjans, G. Behin-Aein, P. Ahmadi, and M. S. Chapman, Phys. Rev. A \textbf{79}, 023401 (2009).

\bibitem{APB} J. Kobayashi, Y. Izumi, K. Enomoto, M. Kumakura, and Y. Takahashi, Appl. Phys. B: Lasers Opt. \textbf{95}, 37 (2009).

\bibitem{HJ} H. Jing, Y. Jiang, and P. Meystre, Phys. Rev. A
\textbf{81}, 031603(R) (2010).

\bibitem{J. C.} J. Cheng,  Phys. Rev. A \textbf{80}, 023608 (2009); J. Cheng, H. Jing, and Y. J.
Yan, $ibid$. \textbf{77}, 061604 (2008).

\bibitem{Inouye} K. Aikawa1, D. Akamatsu, M. Hayashi, K. Oasa, J.
Kobayashi, P. Naidon, T. Kishimoto, M. Ueda, and S. Inouye, Phys.
Rev. Lett. \textbf{105}, 203001 (2010).

\bibitem{Carr} R.V. Krems, W. C. Stwalley, and B. Friedrich, Cold
Molecules: Theory, Experiment, Applications (CRC, Boca Raton, 2009);
L. D. Carr, D. DeMille, R.V. Krems, and J. Ye, New J. Phys.
\textbf{11}, 055049 (2009).

\bibitem{adiabatic elimination} J. Calsamiglia, M. Mackie, and K. A. Suominen, Phys. Rev. Lett. \textbf{87}, 160403 (2001).

\bibitem{dark state} K. Winkler, G. Thalhammer, M. Theis, H. Ritsch, R. Grimm, and J. H. Denschlag, Phys. Rev. Lett. \textbf{95}, 063202 (2005).

\bibitem{Ling} H. Y. Ling, H. Pu, and B. Seaman, Phys. Rev. Lett. \textbf{93}, 250403
(2004); J. Cheng, S. S. Han, and Y. J. Yan, Phys. Rev. A
\textbf{73}, 035601 (2006).

\bibitem{Pu2000} H. Pu and P. Meystre, Phys. Rev. Lett. \textbf{85}, 3987 (2000); L. M. Duan, A. Sorensen, J. I. Cirac, and P. Zoller, $ibid$. \textbf{85}, 3991 (2000).

\bibitem{adiabatic elimination 1} Y. X. Liu, J. Q. You, L. F. Wei, C. P. Sun, and F. Nori, Phys. Rev. Lett. \textbf{95}, 087001 (2005).

\bibitem{three body} B. Borca, J. W. Dunn, V. Kokoouline, and C. H. Greene, Phys. Rev. Lett. \textbf{91}, 070404 (2003); C. P. Search, W. P.
Zhang, and P. Meystre, {\it ibid.} \textbf{92}, 140401 (2004).

\bibitem{Zhang} W. X. Zhang, D. L. Zhou, M.-S. Chang, M. S. Chapman, and L. You, Phys. Rev. Lett. \textbf{95}, 180403 (2005).

\bibitem{Sadler} L. E. Sadler, J. M. Higbie, S. R. Leslie, M. Vengalattore, and D. M. Stamper-Kurn , Nature (London) \textbf{443}, 321 (2006).

\bibitem{hetero} M. Luo, Z. B. Li, and C. G. Bao, Phys. Rev. A {\bf 75}, 043609
(2007).

\bibitem{Daley} A. J. Daley, J. M. Taylor, S. Diehl, M. Baranov, and P. Zoller, Phys. Rev. Lett. \textbf{102}, 040402 (2009).

\bibitem{CPA} C. P. Search and P. Meystre, Phys. Rev. Lett. \textbf{93}, 140405 (2004); C. P. Search, J. M.
Campuzano, and M. Zivkovic, Phys. Rev. A \textbf{80}, 043619 (2009).

\bibitem{Ying} Y. Wu and R. C$\hat{o}$te, Phys. Rev. A \textbf{65},
053603 (2002); H. Jiang, Y. Jiang, and P. Meystre, $ibid$.
\textbf{80}, 063618 (2009).

\bibitem{trimer} S. Knoop, F. Ferlaino, M. Mark, M. Berninger, H. Sch\"obel, H.-C. N\"agerl, and R. Grimm, Nature Phys. \textbf{5}, 227 (2009);
F. Ferlaino, S. Knoop, M. Berninger, W. Harm, J. P. D'Incao, H.-C.
N\"ogerl, and R. Grimm, Phys. Rev. Lett. \textbf{102}, 140401
(2009); H. Jing, J. Cheng, and P. Meystre, $ibid$. Phys. Rev. A
\textbf{77}, 043614 (2008).

\end{thebibliography}
\end{document}